\documentclass[twocolumn,showpacs,preprintnumbers,amsmath,amssymb,showkeys]{revtex4}

\usepackage{graphicx}
\usepackage{dcolumn}
\usepackage{bm}

\begin{document}


\title{Interpretation of the Relaxation Time for the Electrical  Conductivity of Elemental Metals Using the Fluctuation Dissipation Theorem\\}

\author{Tadashi  Hirayama} 
\affiliation{National Astronomical Observatory of Japan, Mitaka, Tokyo, Japan 118-8588}%


\date{\today}

\begin{abstract} 
In an earlier paper we reported that  an empirical formula of the  electrical conductivity $\sigma= e^{2}n_{\rm atom}\tau_{ 0}/(mG)$ agrees with experiments within $\sim20\%$ 
for the most of pure elemental metals at room temperature ranges. Here  $\tau_{0}=\hbar/k_{\rm B}T$ is assumed for `all' metals and
$G$  is summed electron numbers in each atomic shell: e.g. $G$=6 for Cr(3d$^{5}$4s$^{1}$). 
In this paper, we find that the above $\tau_{0}$ can be  deduced if the autocorrelation time of electron fluctuating velocity in a simple fluctuation dissipation theorem is converted  to 
$2\!<\!\!\Delta E\!\!><\!\!\Delta t\!\!>\!\!/k_{\rm B}T$,  and if  this $ <\!\!\Delta E\!\!><\!\!\Delta t\!\!>$ is assumed  equal to $\hbar/2$ of the Heisenberg's minimum uncertainty.
This corresponds to the cloesest approach, or head-on collisions.
Independent from this, we find that $n_{\rm electron}=n_{\rm atom}$ is appropriate for $\sigma$ in most elemental metals . 
In discussing temperature dependence of $\sigma\sim T^{-5}$,  besides use of the Debye temperature unit (${\it \Theta}_{\rm D}$),  the temperature unit of ${\it\Theta}_{\rm P}\equiv \hbar \omega_{pi}/k_{\rm{B}} $ is found to be equally
acceptable. Here $\omega_{\rm pi} $ is the ion plasma frequency, only depending on  $n_{\rm electron}$ unlike somewhat ambiguous $\it{\Theta}_{\rm D}$. 
\end{abstract}

\maketitle
\begin{center}
\bf{1. Introductionn} 
\end{center}

\rm\noindent
In the previous paper \cite{hirayama} (Paper I), we found empirically that absolute experimental values of the electrical
 conductivity $\sigma$ for most of the pure elemental metals
 at room temperature  can be well reproduced ($\sim 20{\%}$) by
\begin{equation}
\sigma=\frac{n_{\rm atom}e^{2} }{mG} \tau_{0},      \label {sigma}
\end{equation}
where the relaxation time $\tau_{0}$ is assumed to be 
\begin{equation}
 \tau_{0}=\frac{\hbar}{k_{\rm B}T} \qquad  \label {tau0}
\end{equation}
for `all' metals. Here $n_{\rm atom}$ is the number density of atom in each metal and $m$ is the true electron mass. The factor
$G$ is integer and is summed numbers of outer electrons in the electron configuration of each atom such as $G=1$ for 
Cu(4s$^{1}$), $G$=1+2=3 for In$^{49}$(5s$^{2}$4p$^{1}$), or $G$=1+4=5 for Nb$^{41}$(3d$^{4}$4s$^{1}$). 
These $G$ values are empirically selected to match the observed $\sigma.$ 
Note that we treat metals of normal geometrical size at ordinary pressure.
  
In contrast to our formulae, the conventional Drude form $\sigma=ne^{2}\tau/m^{*}$ requires the values 
of $Z=n/n_{\rm atom}$ for $n$ (the electron number density), $m^{*}$ (the effective electron mass), and in particular 
$\tau$ (the relaxation time). This is always presented in well-read standard textbooks \cite{aschcroft,ibach,kittel}, 
but unfortunately they do not present conclusive ways how to calculate `absolute values' of $\sigma$. 
As noted in paper I, though Lifshitz and Pitaevskii\cite{lifshitz} and Abrikosov\cite{abrikosov}  suggest 
the form of $\tau_{0}\sim \hbar/k_{\rm B}T$, how accurate their `$\sim$' signs is not clear, nor they treat non-monovalent metals. 
These issues are of fundamental importance, but they are forgotten or ignored as research works in the world of solid state physics today.  

Now, examples of the ratio of $\tau_{0}$ from eq.(2) to $\tau_{\rm obs}$ determined from eq.(1) using the experimental
resistivity and $G=1$ are $\tau_{0}/\tau_{\rm obs}$=0.98(Na), 0.97(K), 0.99(Rb), 1.05(Cu), and 0.95(Au)
at room temperature ranges (hereafter $T_{\rm room}$). Because of this remarkable closeness to unity, it is tempting to assume that there may be an extremely simple
explanation for $\tau_{0}=\hbar/k_{\rm B}T$, which is the major theme of this paper.

For this purpose, first we search for the connection of $\tau_{0}$ to the fluctuation dissipation theorem, 
FDT\cite{kubo_et_al, reichl}.
In order to reproduce $\tau_{0}=\hbar/k_{B}T$, we find that
by converting $H\equiv\frac{1}{2}\int_{0}^{\infty}<mv(0)v(t)>dt$ used in a simple FDT
to $H=<\!\!(\Delta E)^{2}\!\!>^{1/2}<\!\!(\Delta t)^{2}\!\!>^{1/2}$, this $H$ must
take the minimum value of $\hbar/2$ expected from the uncertainty principle.

Secondly, we find that the electron number density $n$ which does not appear in eq.(\ref{sigma}) is satisfied
 by $n=n_{\rm atom}$, i.e. $Z=1$, which is of course only applicable to `conductivity of elemental metals'.

Since eqs.(1) and (2) reproduce the experiments quite well for the most metals at $T_{\rm room}$, we can 
reproduce the observed `absolute values' of $ \rho\equiv 1/\sigma$ at low temperature ranges where $\rho\propto T^{5}$ is
observed, if we employ the Gr\"{u}neisen-Bloch formula.
The Debye temperature $\it{\Theta}_{\rm D}$ itself used here as a temperature unit is 
however as is well-known dependent upon the temperature where it is determined or upon methods of derivation, and thus not unique. 
Then we attempt to use the ion-plasma temperature $\it{\Theta}_{\rm P}$ as an alternative unit, determined from 
$k_{\rm B}\it{\Theta}_{\rm P}=\hbar \omega_{\rm pi}$. Here $\omega_{\rm pi}=(n e^{2}Z^{2}/\epsilon_{\rm 0}M)^{1/2}$ is the ion plasma angular frequency 
where $M$ is the ion mass.
We find that use of the $\it{\Theta}_{\rm P}$-unit for the electrical resistivity and specific heat $C_{\rm v}$ is equally
acceptable, or even better in that it requires only the knowledge of $n$, hence unique.  
\\
\begin{center}
\bf {2. Interpreting Relaxation Time} 
\end{center}

\rm\noindent The classical representation of the fluctuation dissipation theorem (FDT)
for the isotropic electrical conductivity $\sigma$ is as eq.(9.1) in the seminal paper of Kubo \cite{kubo} 
\begin{equation}
\sigma=ne^{2}\tau_{\rm K}/m, \qquad\qquad \\   \label {Kubo-sigma}
\end{equation}
where
\begin{equation} 
\tau_{\rm K}=\frac{m}{k_{B}T}\int_{0}^{\infty}<v(0)v(t)>dt.   \label {tau-Kubo}
\end{equation}
Thus in order to obtain $\tau_{K}=\tau_{\rm 0}$ ($G=1$ is implicitly assumed in eq.(\ref{Kubo-sigma})), we need
\begin{equation}
 H \equiv\frac{\tau_{\rm K}k_{\rm B}T}{2} =\frac{1}{2}\int_{0}^{\infty}<mv(0)v(t)>dt=\frac{\hbar}{2} \label{H}
\end{equation}
Here $v(t)$ is the fluctuating electron velocities parallel to the given electric field.
(Thermal fluctuation is of course expected without electric fields, and is on the order of $(4f(1-f)/N)^{1/2}$, 
in unit of the average energy where $f(\epsilon)$ is the Fermi distribution function and $N$ is the total number of electrons 
in a volume $V$. Thus the fluctuation is occurring around the Fermi energy $E_{\rm F}$ with the width $\approx 2k_{\rm B}T.$ 
See detail in Appendix A) 
  
Besides an obvious relation of the autocorrelation of
 $<\!\!v(0)v(t)\!\!>=<\!\!v(t_{1})v(t+t_{1})\!\!>$ for any $t_{1}$ in the assumed stationary stochastic processes, 
we adopt $H=\:\:<\!\!\frac{1}{2}mv(0)v(t)\Delta t\!\!>$, where $\Delta t$
is chosen to reproduce the value of $H$. This may be not unreasonable if we consider that the 
autocorrelation function rapidly decreases as exp$(-t/\tau)$ for the relaxation processes. We approximate $H=<\frac{1}{2}mv^{2}\Delta t>$,
and interpret (\ref{H}) as expressing
\begin{equation} 
H=<\frac{1}{2}mv^{2}\Delta t>=<\Delta E\Delta t>=\frac{\hbar}{2}.\label{H1} 
\end{equation}
If $mv=\hbar k$ and $v(t)\Delta t=\Delta x$ are used, we still obtain 
\begin{equation} 
H=\frac{1}{2}<\!\!\hbar k(t)\times v(t)\Delta t\!\!>=\frac{1}{2}\hbar<\!\!k\Delta x\!\!>=\frac{\hbar}{2}. \label{H2} 
\end{equation}
While the minimum given in the Heisenberg relation is 
   $<\!\!(\Delta E)^{2}\!\!>^{1/2}<\!\!(\Delta t)^{2}\!\!>^{1/2} \equiv (\Delta E)_{\rm rms}(\Delta t)_{\rm rms}=\hbar/2$,
it was necessary to adopt it to match $\tau_{\rm 0}$.

Reversing the logic, assume that $H=<\!\!\frac{1}{2}mv^{2}\Delta t\!\!>$ takes the minimum uncertainty value of
$\hbar/2$ as suggested in eq. (\ref{H1}) or (\ref{H2}), and assume that eqs.(\ref{Kubo-sigma}) and (\ref{tau-Kubo}) hold, we obtain $\tau_{0}=\tau_{\rm K}$
and reach $\tau_{0}=\hbar/k_{\rm B}T.$ Though it seems rather difficult to escape from this statement, we later in $\S$4 discuss
relation to the case of $\rho=1/\sigma \propto T^{5}$ and physical meaning of eq.(\ref {H1}) at the end of $\S$4 and in
$\S$6 (See also Appendix B). 
\\
\begin{center}
\bf{3. $\bf{n/n_{\rm atom}}$ in Multi-Band Metals}
\end{center}

\rm\noindent
Though eq.(1) needs only values of $n_{\rm atom}$, naturally we wish to know $n$ or $Z\equiv n/n_{\rm atom}$. This $Z$ value is tabulated only
for 20 metals in Kittel's table \cite{kittel}, while the other quantities are tabulated for almost all metals, indicating 
that $Z$ is not easy to assign.
\noindent Let us recall the well-known derivation $\sigma=j_{x}/E_{x}$
for a single band as  preparation for multi-band metals:
e.g. each one of five `d' and one `s' bands in Cr(3d$^{5}$4s${^1})$. We rewrite the current density as
\begin{equation}
j_{x}\!=\!-e\int v_{x}fd\bf k' 
\!=-\it e\int\frac{\hbar k_{x}}{m}(f\!-\!f_{\rm 0})d\bf k', \label{current-density}
\end{equation}
using $\it mv_{x}\!=\!\hbar k_{x}, d\bf k'\rm\!=2\it d\bf k\rm /(2\pi)^{3} and \int \it v_{x} f_{\rm 0}d\bf k'\rm=0$, while
$f_{0}$ and $f $ are  unperturbed and perturbed Fermi distribution function,  respectively. 
Given the electric field $ E_{x}$ in the $x$ direction entering the `steady' Boltzmann equation
\begin{equation}
  \frac{\partial f}{\partial t}-eE_{x}\frac{\partial f}{\hbar\partial  k_{x} }
    \approx -eE_{x} \frac{\partial f_{0}}{\hbar\partial  k_{x} }=-\-\frac{f-f_{0}}{\tau_{0}}, \label{boltz}
\end{equation}
we obtain from eq.(\ref{current-density})
\begin{equation}
\frac{\sigma}{e^{2}\tau_{0}/m}=-\int k_{x}\frac{\partial  f_{0}}{\partial k_{x}}d\bf k'
=\it \int\!f_{\rm 0}\frac{ d\bf k}{\rm 4\pi^{3}}=n,  \label{sigma-2}
\end{equation}
 by eliminating $(f\!-\!f_{0})$ and using $ \bf k$-independent $\tau_{0}$ of eq.(\ref{tau0}).
Because of the factor $\partial  f_{0}/\partial k_{x}$, not only  those bands in which the state density does not extend beyond the Fermi energy
do not contribute to $\sigma $ such as 3d$^{10} $ of Cu as noted in Paper I (see e.g. ref.\cite{ibach} Fig.7.12, left), 
but also only electrons having the  energy of $ (1 \pm 2k_{\rm B}T/E_{\rm F})E_{\rm F} \approx  E_{\rm F} $ are 
contributing to $\sigma$  ($E_{\rm F}$=Fermi energy=2-7eV for $Z$=1). 
Partial integration over $dk_{x}$ within the primitive cell  leads to the second equality. 
 The final equality leads to $n$, where $n=k_{\rm F}^{3}/3\pi^{2}$ and $k_{\rm F}=(2mE_{\rm F})^{1/2}/\hbar$ hold. This
 means that  $\sigma/(e^{2}\tau_{0}/m)$ does not depend upon the state density of electrons  at $E_{F}$, 
nor respective bands, but it is simply equal to $n$. Note that even for the multi-bands discussed below,  
the departure from $d{\bf k}=4\pi k^{2}dk$ (by a multiple factor of 1.2-1.6 for most metals) appears to change
 only the  value of $ E_{\rm F}$ \cite{moruzzi}. The rough derivation above may be sufficient for
the discussion below.

For multi-band metals,
we adopt the total resistivity 
\begin{equation}
\rho=\Sigma\rho_{i},  \label{total-rho}
\end{equation} 
where $\rho_{i} \equiv 1/\sigma_{i}$ as in Aschcroft and Mermin \cite{aschcroft} 
(see eqs.(13.20-22) and discussions followed). 
The average momentum equation can be obtained from eq.(\ref{boltz}) by multiplying 
by $mv_{x}/n $ and integrated over $d\bf v\rm$, where $\partial f/\partial t$
leeds to the inertia term $mdV_{x}/dt$ below.
\begin{equation}
mdV_{x}/dt=-eE_{x}-mV_{x}/\tau_{0}=0. \label{macro-equation-of-motion}
\end{equation}
Obviously electrons are on the average decelerated (or `resisted') by collisions with  ions or by phonon scattering 
$(-mV_{x}/\tau_{0})$.
If there are different kinds of ions such as impurity, we should add up $\Sigma mV_{x}/\tau_{i} 
  \equiv mV_{x}/\tau_{\rm tot}  $ as the summed resistance (as in the Matthiessen rule). 
We assume that the same situation occurs among the different bands in elemental metals which may behave independently or additively. 
This means that for the multi-bands metals $\tau_{0}$ should be replaced by $\tau_{\rm tot}$,
  where $\tau_{\rm tot}$ is given by $1/\tau_{\rm tot}=\Sigma(1/\tau_{i})$.
As a result of $\rho_{i}=m/(ne^{2}\tau_{i})$, the total resistivity is $\rho=\Sigma\rho_{i}=m/(ne^{2}\tau_{\rm tot})$.
Further we assume that every $\tau_{i}$ is equal to $\tau_{0}=\hbar/k_{\rm B}T,$
  as can be surmised by the derivation for the single band. 
Because $G$ is found to be the sum of outer electron numbers in electron configuration, 
which is nothing but numbers of responsible bands,
 $1/\tau_{\rm tot}=\Sigma(1/\tau_{i})=G/\tau_{0}$ results, leading to  $\sigma=e^{2}\tau_{0}n/(mG).$

 Since we found $\sigma=e^{2}\tau_{0}n_{\rm atom}/(mG)$ in eq. (\ref{sigma}) empirically from $G\approx G_{\rm obs}$,
it is compatible with the above only when $Z_{\rm eff}\equiv n/n_{\rm atom}=1$ holds. Namely the `effective' valence electron number $Z_{\rm eff}$
responsible for the electrical conductivity is `unity' for the most of elemental metals!
This is consistent with $G$=1 for Cu(3d$^{10}$2s$^{1})$, while e.g. for the cohesive energy, 
use of $Z$(Cu)=11 may be relevant \cite{moruzzi}. 
It seems thus unnecessary to introduce the effective electron mass $m^{\ast}$ to find $\sigma$ and $\lambda$ 
(thermal conductivity, discussed in Paper I), if not for other purposes.

Summarizing, we assume $\tau_{i}=\tau_{0}$ for each $i$-band and $\rho=\Sigma\rho_{i}\propto\Sigma(1/\tau_{i})
=\Sigma_{i}1/\tau_{0}=G/\tau_{0}$. Since this $\rho=mG/(ne^{2}\tau_{0})$ must be equal to $\rho=mG/(n_{\rm atom}e^{2}\tau_{0})$ from the
empirical eq.(\ref{sigma}), we obtain for $\rho$ of all elemental metals 
\begin{equation}
Z_{\rm eff}\equiv n/n_{\rm atom}=1. \label{zeff}
\end{equation}
Thus e.g. in  Al([Ne]3s$^{2}$3p), $Z_{\rm eff}$ is 1 for $\rho$, different from $Z_{\rm eff}=3$ for the cohesive energy, for example.
Besides, $G$=1 for Al is compatible with the experimental $\rho$ (Paper I). 
Note that our assertion of eqs.({\ref{sigma}) and (\ref{tau0}) holds whether $Z_{\rm eff}=1$ or not.
\\
\begin{center}
\bf{4. Two Temperature Units} 
\end{center}

\rm\noindent To proceed on the theoretical interpretation of $\tau_{\rm K}$, we extend
the resistivity to low temperature ranges, by adopting the Gr\"{u}neisen-Bloch
function $F_{\rm GB}=F_{\rm GB}(T)$ (see e.g. eq(9.62) in ref.\cite{ibach}); $F_{\rm GB}\propto T$ for $T_{\rm red}\gg 1$,
$F_{\rm GB}\propto T^{5}$ for $T_{\rm red}\ll 1$, and $F_{\rm GB}(1)=1.$  
 Here $T_{\rm red}\equiv T/\it{\Theta}_{\rm D}$ and $\it{\Theta}_{\rm D}$ is the Debye temperature ($T_{\rm red}\equiv T_{\rm reduced}).$ 
Thus we extend our formulae of eqs.(1) and (2) to the following, assuming that the Gr\"{u}neisen-Bloch
formula is valid.
\begin{eqnarray}
\rho=\frac{mG}{e^{2}n_{\rm atom}}\frac{1}{\tau_{\alpha}}  \qquad\\ \label{rhoP}
\frac{1}{\tau_{\alpha}}=\frac{1}{\tau_{0}}\frac{F_{\rm GB}(T)/T}{F_{\rm GB}(\it{\Theta}_{\alpha})/\it{\Theta}_{\alpha}}                                                                                                                                                                                                                                                                                                                                                                                                                                                                                                                                                                                                                                                                                                                                                                                                                                                                                                                                                                                                                                                                                                                                                                                                                                                                                                                                                                                                                                                                                                                                                                                                                                                                                                                                                                                                                                                                                                                                                                                                                                                                                                                                                                                                                                                                                                                                                                                                                                                                                                                                                                                                                                                                                                                                                                                                                                                                                                                                                                                                                                                                                                                                                                                                                                                                                                                                                                                                                                                                                                                                                                                                                                                                                                                                                                                                                                                                                                                                                                                                                                                                                                                                                                                                                                                                                                                                                                                                                                                                                                                                                                                                                                                                                                                                                                                                                                                                                                                                                                                                                                                                                                                                                                                                                                                                                                                                                                                                                                                                                                                                                                                                                                                                                                                                                                                                                                                                                                                                                                                                                                                                                                                                                                                                                                                                                                                                                                                                                                                                                                                                                                                                                                            \ \label{tauP} 
\end{eqnarray}%
\noindent Here the subscript $\alpha$ stands either for $\rm D$ or $\rm P$ 
the ion-plasma temperature (`D' for Debye and `P' for plasma). For
the high temperature range where $F_{\rm GB}\propto T$ holds, eq. (15) gives $\tau_{\alpha}=\tau_{0}\propto T^{-1}$ and
we obtain $\rho\propto T.$
\begin{figure}
\begin{center}
\includegraphics[width=8.5cm]{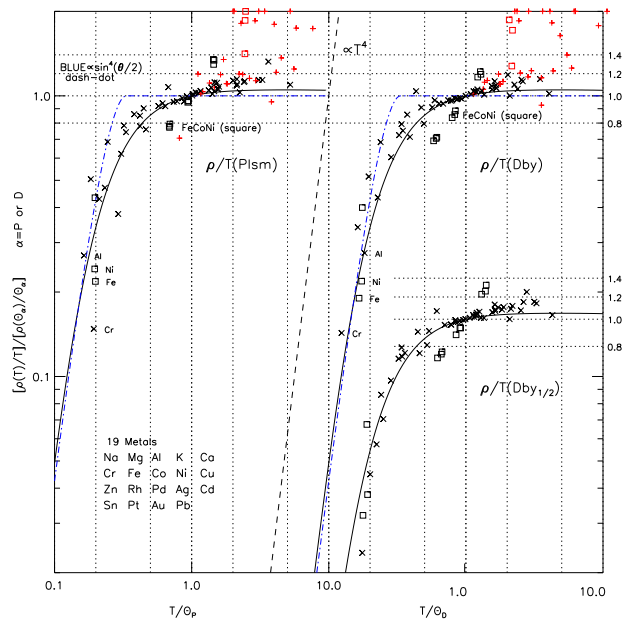}
\caption{Electrical resistivity vs. temperature for 19 metals. The experimental data of 
$(\rho(T)/T)/(\rho(\it{\Theta}_{\alpha})/\it{\Theta}_{\alpha})$ are plotted against
$T/\it{\Theta}_{\alpha}\:$ in black cross sign (left $\alpha$=Plasma and right $\alpha$=Debye) 
(symbolically denoted as $\rho/T$ below respective plots). 
The red plus sign plots are data from $0.5T_{\rm melt}<T<T_{\rm melt}$, the melting point.
Down-shifted plots denoted as Dby$_{1/2}$ use the Debye temperature fitted 
around $T=\it{\Theta}_{\rm D}/\rm 2$ by use of $C_{\rm V}$ adopted from ref.\cite{burns}. 
Three full lines plot the same ratios using Gr\"{u}neisen-Bloch function.
Though Fe, Co and Ni (square) show $\rho\propto T^{1.7}$, there is no discernible difference between $\it{\Theta}_{\rm D}$
and $\it{\Theta}_{\rm P}$ plots. } 
\label{f1}  
\end{center}
\end{figure}

Figure 1 in upper right shows the temperature dependency of the observed electrical resistivity $\rho$ divided by 
temperature, $\rho/T  [\Omega \rm m/ K]$, in unit of $\rho(\it{\Theta}_{\rm D})/\it{\Theta}_{\rm D}$ for 19 metals.
We adpoted the same metals used in Fig. 2 for easy rough comparison. 
Here the Debye temperature $\it{\Theta}_{\rm D}$ is adopted from Kittel \cite{kittel}, which is
empirically determined from the lattice specific heat at constant volume $C_{\rm V}$ in low temperature regions of $C_{\rm V}\propto T^{3}$.
 Because the tabulated temperature steps for $\rho$ \cite{rika} are too sparse, i.e $T$(K)=[78, 273, 573, 973], 
logarithmic interpolation is made to obtain the `observed' $\rho(\it{\Theta}_{\rm D})$. Since we are interested whether
a different unit as compared to the Debye temperature unit can be employed, this kind of roughness is not problematic.
A general trend of proportionality from $T^{4}$ (dashed line in the middle of the figure) to constant, coinciding roughly 
with the $F_{\rm GB}/T_{\rm red}$(full line), is apparent. 
At higher temperature of $T/\it{\Theta}_{\rm D}>\rm 2$, however, the departure from constancy of $\rho/T$ becomes 
appreciable, so that we plotted in red cross signs the observed points in the range $0.5T_{\rm melt}<T<T_{\rm melt}$,
where $T_{\rm melt}$ is the melting points of each metal (factor 0.5 in $0.5T_{\rm melt}$ is arbitrarily chosen).
Discussions hereafter neglect these points.

For reference in the lower right, we show the $\rho/T$-ratio using $\it{\Theta}_{\rm D}$ from ref.\cite{burns} determined at the
temperature near 1/2 of the to-be-determined Debye temperature from the data plots of experimental $C_{\rm V}$.

Because of the well-known ambiguity of adopting the Debye temperature (empirically from specific heat,
acoustic speeds or $\rho$ itself), and its temperature dependency due to the various fitted temperatures as shown above), 
we in this paper attempt to use another temperature $\it{\Theta}_{\rm P}$ expected from
the ion plasma angular frequency $\omega_{\rm pi}$ as defined below:
\begin{equation} 
\it{\Theta}_{\rm P}\equiv \hbar \omega_{\rm pi}/k_{\rm B}, \qquad \omega_{\rm pi} 
\equiv(n e^{\rm 2}Z_{\rm eff}^{\rm 2}/\epsilon_{\rm 0}M)^{\rm 1/2} \label{Tp}
\end{equation}
\noindent
Here $n$ is the electron number density, $\epsilon_{0}$ is the permittivity of vacuum ($1/4\pi$ in c.g.s. Gauss unit) and
$M$[kg] is mass of each atom. The only unknown parameter in determining $\it{\Theta}_{\rm P}$ is the ratio of 
$n/n_{\rm atom}\equiv Z_{\rm eff}$,
which we take unity for all the elemental metals from eq.(\ref{zeff}). Mostly $\it{\Theta}_{\rm D}/\it{\Theta}_{\rm P}$ is 0.9$\sim$ 1.2,
while in alkali, In and Tl it is very close to 1/2 and near 2 in Mo, Rh, W, Os and Ir 
($\it{\Theta}_{\rm D}$ differs among various metals more than $\it{\Theta}_{\rm P}$). 
The dispersion relation is $\omega=\omega_{\rm pi}k/(k_{\rm TF}^{2}+k^{2})^{1/2}.$
Here $k_{\rm TF}(=[e^{2}D(E_{\rm F})/\epsilon_{0}]^{1/2}=(4k_{\rm F}/\pi a_{0})^{1/2}\leq n^{1/3})$ is the Thomas-Fermi shielding wave number, 
where $D(E_{\rm F})$ is the electron state density at the Fermi energy, $k_{\rm F}$ is its wave number and $a_{0}$ is 
the Bohr radius (e.g. ref.\cite{aschcroft} in eqs.(17.55) and (26.4)).

Figure 1 left plots $[\rho(T)/T]/[\rho(\it{\Theta}_{\rm P})/\it{\Theta}_{\rm P}]$ against $T/\it{\Theta}_{\rm P}$ as in the Debye-plotting. 
Both plots either by use of Debye or Plasma temperature unit are found quite similar, indicating that to adopt $\it{\Theta}_{\rm P}$ is not out of question, 
irrespective of theoretical reasoning. Full black lines are all $F_{\rm GB}$ functions, passing through the point of (1.0,1.0), and of course
it is identical to the ones which used $\it{\Theta}_{\rm D}$ unit.
Thus if we adopt eqs.(14) and (\ref{tauP}) for $\alpha$=P, namely adopt the temperature expected from 
the ion-plasma frequency as in eq.(\ref{Tp}), most of the `absolute' values of resistivity of elemental metals can be well reproduced without introducing
empirical parameters other than $G$ in a temperature range somewhat below the melting points at ordinary pressure. 
Eq.(14) with $\alpha=P$ might be used for further study, because, as is well-known,
not only Fe, Co, and Ni depart appreciably from  this in $T_{\rm red}=1-2$, but also below a few  Kelvin 
many metals \cite{mendelssohn,levy} show
 $\rho$=const+const$\times T^{j}$ including Na, Cu, and Ag, where $j\approx$ 2 (ref.\cite{psj} lists 10 such metals).  

In order to see the theoretical temperature dependency of $\tau_{\alpha}$ or $\rho$, we show below a simplified temperature
dependency of the collision time, denoted as $\tau_{\theta}$. 
\begin{equation}
\frac{1}{\tau_{\theta}}=\frac{1}{\tau_{0}}\int_{0}^{\theta}(1-\cos\theta)\cdot 2\pi\sin\theta \frac{d\theta}{4\pi}
  =\frac{1}{\tau_{0}}\sin^{4}(\frac{\theta}{\rm 2}) \label{sin4}
\end{equation}

\noindent for $\theta\le \pi$ and $\tau_{\theta}=\tau_{0}$ for $\theta>\pi.$ 
This is equivalent to assume $W(\theta)=1/\tau_{0}$ in eq.(3.11) of ref.\cite{abrikosov}, or in eq.(16.32) of ref.\cite{aschcroft}.
Here $1-\cos \theta$ is the weight factor to the direction of the given electric fields and $2\pi \sin \theta d\theta/4\pi$
is the fraction of the differential solid angle.
Further we simplify that $\theta=T/\it{\Theta}_{\rm D}$ instead of the standard $\sin(\theta/2)=q/2k_{\rm F}$ where $q$ is 
the phonon wave number and $k_{\rm F}$ is that for electrons at the Fermi energy. 
Blue dash-dot lines in Fig.1 left and right plot $Y(\rm ordinate)=sin^{4}\it(T_{\rm red} $/2)) against 
$X=T_{\rm red} /(\rm{3}\pi/2)$ for $T_{\rm red} \leq \pi$, while for $\theta>\pi$/2, $Y$=1 is kept. 
These tend to $\rho/T\propto T^{4}$ for the low temperature, coinciding with
$\rho\propto T^{5}$, and for the high temperature we obtain $\rho \propto T$, which are rather satisfactory.
This is found to be the upper envelope of the observed data points near $X(\rm abscissa)\approx 0.3.$ 
The much simplified eq.(\ref{sin4}) is, if any merit in it, to show the $\theta$ dependency clearly. 
At the same time $X=T_{\rm red}$
and $Y=\sin^{4}(T_{\rm red} \times 3\pi/2),$ and $Y=1$ for $X>1/3$ give an identical curve,
which is in some cases convenient since the same $X$ value as the Gr$\rm \ddot{u}$neisen curve is used;
e.g the maximum ratio between two curves is 1.6 at $T_{\rm red}$=0.2.

\begin{figure}
\begin{center}
\includegraphics[width=8.5cm]{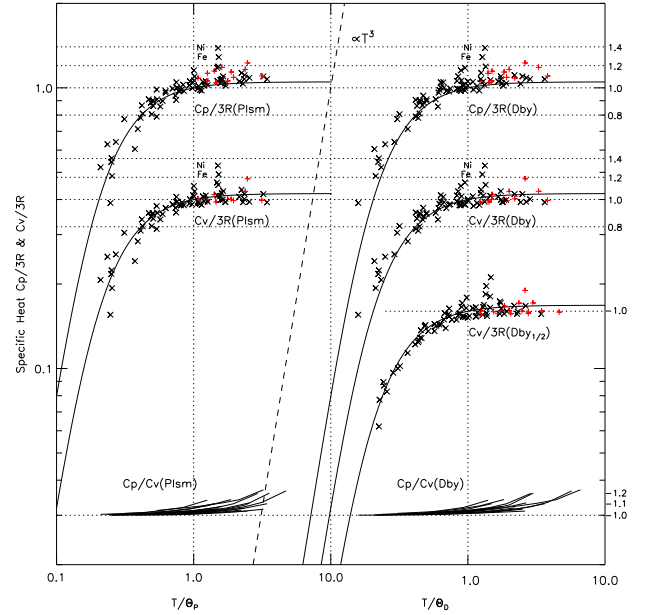}
\caption{Specific heat (usual one from lattice) vs. temperature. Experimental values of $C_{\rm P}/(3R)$ and $C_{\rm V}/(3R)$ are plotted
against $T/T_{\alpha}$ in black cross marks ($\alpha=P$, left and $\alpha=D$, right). Red plus signs are those at temperatures
 $0.5T_{\rm melt}T<T_{\rm melt}$ as in Fig.1.
Upper 5-full lines are all the same Debye functions. Two bottom figures (left and right) plot $C_{\rm P}/C_{\rm V}$ 
for each metal used for conversion from $C_{\rm P}$ to $C_{\rm V}$.}
\label{fig2}  
\end{center}
\end{figure}

Figure 2 left plots the observed $C_{\rm P}/(3R)$ from ref.\cite{rika} (upper) and $C_{\rm V}/(3R)$ calculated from
$C_{\rm P}/(3R)$ (lower, discussed below) against $T/\it{\Theta}_{\rm P}$, and in 
the right, similar plots against $T/\it{\Theta}_{\rm D}$ both for 19 metals. Hrer $R$ is the gas constant.
The temperature step is again sparse; [100, 200, 298.15, 400, 600]\,K. If we compare the left Cv/3R(plsm), 
meaning plasma, and right Cv/3R(Dby), there is no large difference between the two,
and the Cv/3R(plsm) by the use of the ion plasma temperature shows smaller scatter from the Debye function (full lines).
Thus use of the plasma temperature as the temperature unit, which needs no adjustments from experiments, 
is not restricted to the resistivity as in Fig.1 such that eqs.(14) and (\ref{tauP}) with $\alpha=P$ may be allowed to be
one possible standard.
 
Below are some remarks. The bottom curves are $C_{\rm p}/C_{\rm V}$ from 
$C_{\rm P}(\rm{JK^{-1}mol^{-1}})$=$C_{\rm V}+(3\alpha_{\rm L})^{2}TV_{\rm m}/\chi$ for each metal. Here $\alpha_{\rm L}=\alpha_{\rm L}(T)$ 
is the linear thermal expansion coefficient \cite{rika}, $\chi$ is the isothermal compressibility \cite{kittel}, and
$V_{\rm m}$ is the volume/mol(=atomic weight/metal density).
The down-shifted plot denoted as $C_{\rm V}/3R$ (Dby$_{1/2}$) as in Fig. \!1 shows naturally smaller scatter than the plot
denoted as Cv(Dby) from the Debye temperature
determined at much lower temperature. Red plus plots are for $0.5T_{\rm melt}\!<\!T<\!T_{\rm melt}$ as before. Full lines are all
the same exact Debye functions, which again passes (1,1)-point and the limiting
value is 1.05014=1/0.95225. Dashed line is the limiting $C_{\rm V}\propto T^{3}$. 

If the same correction factor $C_{\rm V}/C_{\rm P}$ is tentatively applied
to $(\rho/T)_{\rm ratio}$ in Fig.1, the corrected plot becomes closer to $\rho/T$=constant for $T/T_{\alpha}\ge 1$. 
It is clear that some kind of small correction to $\rho/T$ is needed, because the 
electron density will decrease due to larger separation between atoms when temperature increases.

The first conclusion of this section is that use of $\it{\Theta}_{\rm P}$ from the ion plasma frequency as a temperature unit
gives reasonable temperature dependence for both the resistivity and specific heat, and it is unique as compared to 
$\it{\Theta}_{\rm D}$, such that eqs.(14) and (15) for $\alpha$=P may be more appropriate.
Now the applicable range of  $\tau_{\rm K}$ is restricted to e.g. $\rho(T)/T\approx \rho(\it{\Theta}_{\rm P})/\it{\Theta}_{\rm P}$,
meaning the deflection angle  be $\theta>\pi/2$ as seen particularly in the $\sin^{4}(\theta/2)$-plot of Fig.1. 
The latter implies the close encounter, or  a small impact parameter,
and suggests, as the second conclusion of this section, that the minimum uncertainty of $\hbar/2$ is maintained
for $\tau_{\rm K}$ and hence $\tau_{0}$ may well be due to the close collisions. These arguments could have been presented by
employing the Debye temperature unit, but now by the use of $\it{\Theta}_{\rm P}$ these become less ambiguous.

We note a similarity between the present close collision and the electron in the first Bohr orbital in atomic hydrogen, where the quantized angular
momentum $mvr=\hbar$ is conserved in the circular and closest orbital. In the present case of 
the elastic collision in a hyperbolic orbital, the angular momentum is still conserved, and its minimum absolute value is 
$mvr=\hbar kr=\hbar/2$. 
\\
\begin{center}
\bf{5. Electron Specific Heat} 
\end{center}

\rm\noindent
The electron specific heat, discernible only below a few Kelvin or so, is given by 
$C_{\rm Ve}=\pi^{2}D(E_{\rm F})k_{\rm B}^{2}T/3$, where $D(E_{\rm F})$
is the number of states per energy per atom. For the free electron model 
$D(E_{\rm F})=3/(2E_{\rm F})$ holds, leading to $C_{\rm Ve}=\pi^{2}k_{\rm B}^2T/2E_{\rm F}$.  
The ratio $C_{\rm Ve,obs}/C_{\rm Ve,theo}$ is conventionally expressed by $m^{*}/m\,(m^{*}$=the thermal effective electron mass), 
and $m{*}/m$ amounts to 10 or more for transition metals if the free electron model is used for the theoretical 
$C_{\rm Ve,theo}$ (see e.g. ref.\cite{ibach}, table 6.2). 

However if we adopt $D(E_{\rm F})$ from the extensive calculation of Moruzzi $\it{et\;al}$.\cite{moruzzi} and 
$C_{\rm Ve,obs}$ from ref.\cite{kittel}, we find on the average $m^{*}/m=1.48\pm$0.89 for 30 metals down to the period starting from Rb to In.
If we exclude six metals showing $m^{*}/m>2.0$, the average becomes $m^{*}/m=1.11\pm 0.29$, which is quite satisfactory. 

Values of $m/m^{*}$ for excluded six metals are 2.6(Sc), 2.4(V), 2.5(Mn), 4.9(Sr), 3.1(Y), and 2.4(Nb). 
For example, Sr shows a sharp drop in $D(E)$ near $E_{\rm F}$\cite{moruzzi}, suggesting that $D(E_{\rm F})=0.31$ could have been much 
larger, and accordingly $m/m^{*}$ may become closer to unity. 
A similar situation is seen in Y ($D(E_{\rm F})=1.41)$, which has a steep peak of $D(E)$ just below $E_{\rm F}$. 

Thus we can safely say that the
theoretical understanding of $C_{\rm Ve}$ is rather satisfactory for the elemental metals (except perhaps for semi-metals not included
in Moruzzi $\it{et\;al}$.), so that we need not concern it 
when discussing other physical quantities.
\\
\begin{center}
\bf{6. Discussion and Conclusion} 
\end{center}
\rm\noindent
We found that the empirically well established $\tau_{0}=\hbar/k_{\rm B}T$ for the room temperature range in elemental
metals (Paper I) can be deduced in the following way. We take conversion
\begin{equation}
\int_{0}^{\infty}\!<\!\frac{m}{2}v(0)v(t)\!>dt\!=<\!\frac{mv^{2}}{2}\Delta t\!>=<\!\Delta E\Delta t\!> 
\end{equation}
(see Appendix A for the second equality), and assume 
\begin {equation}
<\!\Delta E\Delta t\!>=\hbar/2, \label{Et_eq_h}
\end{equation}
then we obtain
\begin{equation}  
\tau_{\rm K}\equiv\frac{m}{k_{B}T}\int_{0}^{\infty}<v(0)v(t)>dt=\frac{\hbar}{k_{\rm B}T}=\tau_{0}.   
\end{equation}

Further simplified $\tau_{0}$-derivation is to assume from the outset that the relaxation time $\tau_{0}$ in the steady Boltzmann
equation is equal to  
the autocorrelation time of the normalized fluctuating thermal velocity $\tau_{\rm K}\,$
as in eq.(B1). Namely
\begin{equation}
-eE_{x} \frac{\partial f_{0}}{\hbar\partial  k_{x} }=-\frac{f-f_{0}}{\tau_{\rm k}}, \label{boltz1}
\end{equation}
where
\begin{equation}
\tau_{\rm K}\!=\!\int_{0}^{\infty }\!\!<\!\frac{v(0)}{v_{T}}\frac{v(t)}{v_{T}}\!>\!dt\!.
\label{tauk-c1}
\end{equation}

 This might be regarded as a basis of derivation of the simple classical FDT, if so wished. Then we evaluate 
$\tau_{\rm K}$, assuming eq.(19) and using eq.(18). In this case $\tau_{0}$ entering the steady Boltzmann equation [\ref{boltz}]
should better be called `correlation time' rather than `relaxation time'. At the same time the once discarded `collision time',
if not by all people, be retained because without notion of `collisions', $\tau_{0}=\hbar/k_{\rm B}T$ cannot be understood in closest collisions.

For the wider temperature ranges, we can well reproduce the observed
`absolute' values of electrical resistivity $\rho$ by eqs.(14) and  (\ref{tauP}), exception being $T$ higher than a fraction of $T_{\rm melt}$. Here the latter assumes the 
Gr\"{u}neisen-Bloch function, where only the relative $\rho$ values can be given as eq.(9.62) of ref. \cite{ibach}. 

Further, we find that $n_{\rm atom}$ used in eqs.(1) and (14) is equal to $n$, i.e. $Z_{\rm eff}=1$ for $\rho$ 
and for the thermal conductivity too (Paper I), though only in elemental metals, and not necessarily applicable to other
applications (a short summary above eq.(13)). 

If we accept $Z_{\rm eff}$=1, the temperature unit from the ion-plasma frequency, 
$k_{\rm B}\it{\Theta}_{\rm P}\equiv \hbar \omega_{\rm ip}$, becomes
useful, in that the unique temperature value only dependent upon the electron density $n$ can be defined such that
the observed $\rho/T$ and $C_{\rm v}$ can be represented quite 
similarly or slightly better than use of $\it{\Theta}_{\rm D}$ unit (Figs.1-2).
Regarding the excitation of ion-plasmons in eq.(\ref{Tp}) of $\hbar\omega_{\rm pi}=k_{\rm B}\it{\Theta}_{\rm P}$, one might think of 
equilibration with the ubiquitous blackbody radiation inside the metal, though one needs a careful study.

Eq.(\ref{Et_eq_h}), $<\!\!\Delta E\Delta t\!\!>=\hbar/2$, is applicable to the large angle scattering, implying close collisions with small impact parameters.
We mentioned at the end of $\S$4 the similarity of the quantized angular momentum $mvr=\hbar$
 in the first Bohr orbital. 
Another comparison is to use a harmonic oscillator (one-dimensional, 1-D) where $\Delta E\Delta t=\hbar/2$ is realized
in the closest oscillation with a Gaussian wave function,
and to consider collisions with the scattering angle of $\theta=\pi$, i.e. 1-D head-on collisions (zero impact parameters).
Difference between the two is bounded or unbounded motion, but with the same closest approach, giving the same
 $<\!\!\Delta E\Delta t\!\!>=\hbar/2$. 
\appendix
\label{App}
\section{\bf{Fluctuation Amplitudes}}

\rm\noindent In order to find the amplitude of energy fluctuations, we use the fluctuation of electron numbers $\Delta N$ in a volume $V$ 
as given in Landau and Lifshitz \cite{landau}, eq.(113.6)
\begin{equation}
\label{App1}
<(\Delta N)^{2}>=G_{\rm N}f\times(1-f),
\end{equation}

\noindent which is derived from partial derivative of the Fermi distribution function $f=1/\left\{\exp[(\epsilon-\mu)/k_{B}T]+1\right\}$ with respect
to the chemical potential $\mu$. This $\mu$ is practically the same as the Fermi energy $E_{\rm F}.$ Here $G_{\rm N}$ is a scalar of states number
given by 
\begin{equation}
\label{App2}
G_{\rm N}=d\left[2\frac{4\pi}{3}(\frac{\bf{p}}{h})^{3}\right]V=\frac{k^{3}}{\pi^{2}}\frac{dk}{k}V
=\frac{3}{2}\frac{d\epsilon}{E_{\rm F}}N, 
\end{equation}

\noindent where use is made of $\bf{p}=\hbar \bf{k}$, $n=3\pi^{2}k_{\rm F}^{3}, dk/k_{\rm F}=d\epsilon/2E_{\rm F}$, and $N=nV$
($G_{\rm N}=DNd\epsilon$, where $D$ is the conventional state density).  
Multiplying eq.(\ref{App1}) by $\epsilon^{2}$ and using $<\!\!(\Delta E)^{2}\!\!>\equiv<\!\!(\epsilon\Delta N)^{2}\!\!>$, 
we find
\begin{equation}
\frac{\sqrt{<\!(\Delta E)^{2}\!>}}{<E>}=\left[\frac{3f(1-f)}{2N}\frac{d\epsilon}{E_{\rm F}}\right]^{1/2}\propto\frac{1}{\sqrt{N}} \label{rel_fluct}
\end{equation}

The relative fluctuation eq.(\ref{rel_fluct}) is dependent upon sizes $(\propto 1/\sqrt{N})$ as in the Boltzmann distribution
and dependent upon the width of energy spectrum $d\epsilon$.
Besides, it is appreciable only near $\epsilon\approx E_{\rm F}$ (max[$f(1-f)$]=1/4) such that 
below $\epsilon\simeq E_{\rm F}-2k_{\rm B}T$, $f$ is saturated at unity so that there is no fluctuations in $\Delta N$ 
nor $\Delta E$ there. Thus 
\begin{equation}
\sqrt{<\!(\Delta E)^{2}\!>}=<mv^{2}/2>
\end{equation}
\noindent 
should hold, where $mv^{2}/2$ is the thermal energy measured from the center value of $\epsilon=E_{\rm F}$.
Note that if we adopt $f(1-f)d\epsilon \approx d\epsilon \approx k_{\rm B}T$, this becomes identical to 
the problem-answer at the end of $\S$113 in ref.\cite{landau}.\\

\section{\bf{Correlation Time in FDT}}

\noindent As emphasized by Kubo \cite{kubo} in p.580 , unless one introduces some physical models the FDT does not give 
useful answers, even though it may be correct.
This is shown in the `simplest example of the independently moving charged particles in the classical system' (p.585), i.e. 
eq.(4). 
Utilizing  a normalized velocity function $v(t)/v_{T}$ where $v_{\rm T}=(k_{\rm B}T/m)^{1/2}$
and introducing the normalized autocorrelation function $C$, eq.(4) becomes
\begin{equation}
\tau_{\rm K}\!=\!\int_{0}^{\infty }\!\!<\!\frac{v(0)}{v_{T}}\frac{v(t)}{v_{T}}\!>\!dt\!
=\!\int_{0}^{\infty }\!\!C(t)dt\!
=\!\tau_{\rm cor}.   \label{tauk-c}
\end{equation}
This is almost a definition of correlation time as seen in  
$C(t)=C(0)\rm exp\it(-t/\tau_{\rm cor})$. But if it is so, $\tau_{0}=\tau_{\rm K}$ might be used directly in the Boltzmann equation
eq.(\ref{boltz}). 

\end{document}